\documentclass[12p]{article}
 \usepackage{dcolumn}

\usepackage{latexsym}
\usepackage{cite}
\usepackage{color}
\usepackage{amsmath}
\usepackage{epsfig}
\usepackage{hyperref}
\newcommand{\ortala}[1]{\begin{center}#1\end{center}}

\newcommand{\sandd}[1]{\left\langle #1\right\rangle}

\newcommand{\summ}[3]{{{\underset{#1 }{\overset{#2}{\displaystyle\sum}}}#3}}
\newcommand{\prodd}[3]{{{\underset{#1
}{\overset{#2}{\displaystyle\prod}}}#3}}

\newcommand{\re}[1]{(\ref{#1})}

\newcommand{\eq}[2]{\begin{equation}\label{#1}  #2\end{equation}}

\newcommand{\paran}[1]{\left(#1\right)}

\newcommand{\sch}[1]{Schrodinger}
\newcommand{\komb}[2]{\paran{\begin{array}{c} #1 \\ #2 \end{array}}}

\linethickness{0.6pt}

\begin{document}

\ortala{\textbf{Hysteresis behaviors of the crystal field diluted  general spin-S Ising model}}

\ortala{\"Umit Ak\i nc\i\footnote{umit.akinci@deu.edu.tr}}
\ortala{\textit{Department of Physics, Dokuz Eyl\"ul University,
TR-35160 Izmir, Turkey}}

\section{Abstract}\label{abstract}

Hysteresis characteristics of the crystal field diluted general Spin-S ($S>1$) Blume-Capel model have 
been studied within the effective field
approximation. Particular emphasis has been paid on the large negative valued crystal field and low temperature region 
and it has been demonstrated for this region that, rising dilution of the crystal field results in decreasing number of windowed hysteresis loops. The evolution
of the multiple hysteresis loop with the dilution of the crystal field has been investigated and physical mechanism behind this evolution has been given.

\section{Introduction}\label{introduction}

S-1 Blume Capel (BC) model \cite{ref1,ref2} is one of the most extensively studied models, after S-1/2 Ising model \cite{ref3}, which is the basic model of magnetism in statistical physics. One of the extensions of the model is, S-1 BC model with random crystal field distribution, which mimics $He^3-He^4$ mixtures in a random media, i.e. aerogel. The special case of  this class of models is S-1 BC model with quenched crystal field dilution, which means that, $p$ percentage of the lattice sites are under the influence of the crystal field $D$ and there is no crystal field effect on the remainig sites.

The phase diagrams and thermodynamical properties of the S-1 BC model has been widely studied with several methods, such as the renormalization group (RG) technique \cite{ref4},  Monte Carlo (MC) simulations \cite{ref5}, cluster variational method (CVM) \cite{ref6,ref7}, finite cluster approximation (FCA) \cite{ref8,ref9}, pair approximation (PA) \cite{ref10}, replica method \cite{ref11}, mean field approximation (MFA) \cite{ref12,ref13,ref14,ref15,ref16,ref17} and effective field theory (EFT) \cite{ref18,ref19,ref20,ref21,ref22,ref23}.
Besides, some other geometries have been investigated, such as Bethe lattice \cite{ref24}, thin film \cite{ref25,ref26}, nanowire \cite{ref27}, semi infinite geometry \cite{ref28}.
 
As in the case of Spin-S BC model without any random distributions, there is a downward trend with the rising spin value in the theoretical literature. S-3/2 random field distribution has been investigated within the MFA \cite{ref29}, EFT \cite{ref30} and RG \cite{ref31}.
Also, S-2 BC model with  random field distribution has been investigated within the MFA \cite{ref32} and EFT \cite{ref33}. General investigation of the 
Spin -S BC model has been studied within the PA \cite{ref34}. Besides, related literature contains also mixed spin random crystal field diluted BC models such as mixed S-1/2 and S-3/2 model \cite{ref35,ref36}, mixed S-1 and S-3/2 model \cite{ref37,ref38} and S-2 and S-5/2 model \cite{ref39}.

Very recently, it has been shown by the author that, crystal field diluted $S-1$ 
BC model could exhibit double and triple hysteresis behaviors at large negative 
values of the crystal field and the physical mechanisms behind  these  
behaviors have been explained \cite{ref40}. This work can be considered as the continuation of that work.
The aim of this work is to obtain general results about the hysteresis behaviors of the crystal field diluted Spin-S BC model. The formulation has been constructed within the EFT \cite{ref41}. The paper is organized as follows: In Sec. \ref{model} we briefly
present the model and  formulation. The results and discussions are
presented in Sec. \ref{results}, and finally Sec. \ref{conclusion}
contains our conclusions.

\section{Model and Formulation}\label{model}

The Hamiltonian of the spin-S BC model with uniform longitudinal
magnetic field is given by
\eq{denk1}{\mathcal{H}=-J\summ{<i,j>}{}{s_is_j}-\summ{i}{}{D_is_i^2}-H\summ{i}{}{
s_i},}
where $s_i$ is the $z$ component of the spin operator and it takes number of 
 $2S+1$ different values
such as $s_i=-S,-S+1,\ldots S-1,S$, $J>0$ is the ferromagnetic
exchange interaction between the nearest neighbor spins, $D_i$ is
the crystal field (single ion anisotropy) at a site $i$, $H$ is the
external longitudinal magnetic field. The first summation in Eq.
\re{denk1} is over the nearest-neighbor pairs of spins and the other
summations are over all the  lattice sites.
Crystal field dilution means that diluting homogenous distributed crystal field $D$ from the system,
i.e. crystal field distributed on lattice sites according to the distribution function

\eq{denk2}{
P\paran{D_i}=p\delta\paran{D_i-D}+\paran{1-p}\delta\paran{D_i},
} where $\delta$ is the delta fuction and p is a real number which can take values between zero and one.

We can construct the EFT equations by starting with generalized Callen-Suzuki 
\cite{ref42} identities, which are generalized versions of the identites for the 
$S-1/2$ system \cite{ref43,ref44} and given as

\eq{denk3}{\sandd{s_0^i}=\sandd{\frac{Tr_0 s_0^i
\exp{\paran{-\beta \mathcal{H}_0}}}{Tr_0\exp{\paran{-\beta
\mathcal{H}_0}}}},} where, $i=1,2,\ldots, 2S$ for the spin-S Ising system (corresponding to the dipol moment, quadrupol moment, octupol moment etc.) and 
$Tr_0$ is the partial trace over the site
$0$. Here, $\mathcal{H}_0$ denotes all interactions of the spin that 
belongs to the site $0$ and it has two parts as spin-spin interactions (denoted 
as $E_0$) and interactions with the fields (crystal field and magnetic field). 
From the Hamiltonian of the system represented  by Eq. \re{denk1},  $\mathcal{H}_0$ is 
given by 

\eq{denk4}{\mathcal{H}_0=s_0\paran{E_0+H}+s_0^2D_0, \quad 
E_0=\summ{\delta=1}{z}{s_\delta},
} where $z$ is the number of nearest neighbor interactions i.e. coordination number and 
$s_\delta$ is the nearest neighbor of the spin located at site $0$.  
Evaluating Eq. \re{denk3} by using Eq. \re{denk4} we can obtain closed form expressions as
\eq{denk5}{\sandd{s_0^i}=\sandd{F_i\paran{E_0,H,D_0}}.
} For including the effect of the crystal field distribution in the formulation, the rhs of Eq. \re{denk5} has to be integrated  over the probability distribution  given by Eq. \re{denk2}, i.e.  

\eq{denk6}{\sandd{s_0^i}=\sandd{G_i\paran{E_0,H,D}}, 
} where 

\eq{denk7}{G_i\paran{E_0,H,D}=p F_i\paran{E_0,H,D} +\paran{1-p} F_i\paran{E_0,H,0}.
} 

Eq. \re{denk6} can be handled by differential operator technique \cite{ref45}. 
 If $\nabla$ is the differential operator with respect to $x$ and  $a$ is any 
constant, then 
\eq{denk8}{\exp{\paran{a\nabla}}G\paran{x}=G\paran{x+a},} 
defines the effect of the exponential differential  operator on any function $G(x)$. With the help of this expression, Eq. \re{denk6}
can be written as 
\eq{denk9}{
\sandd{s_0^i}=\sandd{\exp{\paran{E_0\nabla}}}G_i\paran{x,H,D}.
}

Exponential differential operator in Eq. \re{denk9} can be treated by approximated van der 
Waerden identities which were proposed for the higher spin 
problems \cite{ref46} and it is given as
\eq{denk10}{
\exp{\paran{as_k}}=\cosh{\paran{a\eta}}+\frac{s_k}{\eta} \sinh{\paran{a\eta}},
} where $\eta^2=\sandd{s_k^2}$. This approximation reduces the number of $2S$ equations given by Eq. \re{denk6} to two equations, since this approximation equates $s_k^{2n}$ to $\sandd{s_k^2}^n$ and $s_k^{2n+1}$ to $s_k\sandd{s_k^2}^n$ and they are given by

\eq{denk11}{m=\sandd{s_0}=\sandd{\prodd{k=1}{z}{\left[\cosh{\paran{J\eta\nabla}}
+\frac{s_k}{\eta} \sinh{\paran{J\eta\nabla}}\right]}}G_1\paran{x,H,D},
}
\eq{denk12}{\eta^2=\sandd{s_0^2}=\sandd{\prodd{k=1}{z}{\left[\cosh{\paran{
J\eta\nabla}}+\frac{s_k}{\eta} 
\sinh{\paran{J\eta\nabla}}\right]}}G_2\paran{x,H,D}.
}

Decoupling the terms $s_k$ and $s_k^2$ in the expanded form of Eqs. \re{denk11} 
and \re{denk12}, and using the translationally invariance property of the 
lattice we arrive the equations:

\eq{denk13}{m=\left[\cosh{\paran{J\eta\nabla}}+\frac{m}{\eta} 
\sinh{\paran{J\eta\nabla}}\right]^zG_1\paran{x,H,D},
}
\eq{denk14}{\eta^2=\left[\cosh{\paran{J\eta\nabla}}+\frac{m}{\eta} 
\sinh{\paran{J\eta\nabla}}\right]^zG_2\paran{x,H,D}
.} The explicit forms of the functions in Eqs. \re{denk13} and \re{denk14} can 
be obtained by performing partial trace operations in Eq. \re{denk3} for 
arbitrary $S$ and by using Eq. \re{denk7}, hence they are given as
\eq{denk15}{G_1\paran{x,H,D}=\frac{\summ{k=-S}{S}{}k\exp{\paran{\beta D 
k^2}\sinh{\left[\beta k\paran{x+H}\right]}}}{\summ{k=-S}{S}{}\exp{\paran{\beta D 
k^2}\cosh{\left[\beta k\paran{x+H}\right]}}}
+
\frac{\summ{k=-S}{S}{}k\sinh{\left[\beta k\paran{x+H}\right]}}{\summ{k=-S}{S}{}\cosh{\left[\beta k\paran{x+H}\right]}}}

\eq{denk16}{G_2\paran{x,H,D}=\frac{\summ{k=-S}{S}{}k^2\exp{\paran{\beta D 
k^2}\cosh{\left[\beta k\paran{x+H}\right]}}}{\summ{k=-S}{S}{}\exp{\paran{\beta D 
k^2}\cosh{\left[\beta k\paran{x+H}\right]}}}
+
\frac{\summ{k=-S}{S}{}k^2\cosh{\left[\beta k\paran{x+H}\right]}}{\summ{k=-S}{S}{}\cosh{\left[\beta k\paran{x+H}\right]}}
}

With the help of Binomial expansion and by using Eq. \re{denk8}, Eqs. \re{denk13} and \re{denk14} can be written as  

\eq{denk17}{ m=\summ{n=0}{z}{}{}\komb{z}{n}\frac{m^n}{\eta^n}A_{n}^{(z)},
} 
\eq{denk18}{ \eta^2=\summ{n=0}{z}{}{}\komb{z}{n}\frac{m^n}{\eta^n}B_{n}^{(z)},
} and the coefficients are given by 

\eq{denk19}{
A_{n}^{(z)}=\frac{1}{2^z}\summ{p=0}{z-n}{}\summ{q=0}{n}{}\komb{z-n}{p}\komb{n}{q
}
(-1)^{q}
G_1[\eta J (z-2p-2q),H,D], }
\eq{denk20}{
B_{n}^{(z)}=\frac{1}{2^z}\summ{p=0}{z-n}{}\summ{q=0}{n}{}\komb{z-n}{p}\komb{n}{q
}
(-1)^{q}
G_2[\eta J (z-2p-2q),H,D]. }

By solving the system of nonlinear equations given by  Eqs.
\re{denk17} and \re{denk18}  and using the coefficients given by Eqs.  \re{denk19} 
and \re{denk20}  we
get EFT results for the spin-S BC model. Linearization  of Eqs. 
\re{denk17} and \re{denk18} in $m$ will yield equation system for the second order 
critical temperature.    Detailed investigation of the formulation used here can 
be found  in review article \cite{ref41}.

\section{Results and Discussion}\label{results}

Our inspection is on simple cubic lattice (i.e. $z=6$) within this work. Numerical calculations are performed for the following  scaled (dimensionless) quantities

\eq{denk21}{ d=D/J,t=k_BT/J,h=H/J. }

The hysteresis loops can be obtained for a given parameter set
($p,d,t$) by calculating the $m$ according to the procedure given
above, and by  sweeping the longitudinal magnetic field from $-h_0$
to $h_0$ and after then in reverse direction, i.e. $h_0\rightarrow - h_0$. 

\subsection{Phase diagrams}

\begin{figure}[h]
\epsfig{file=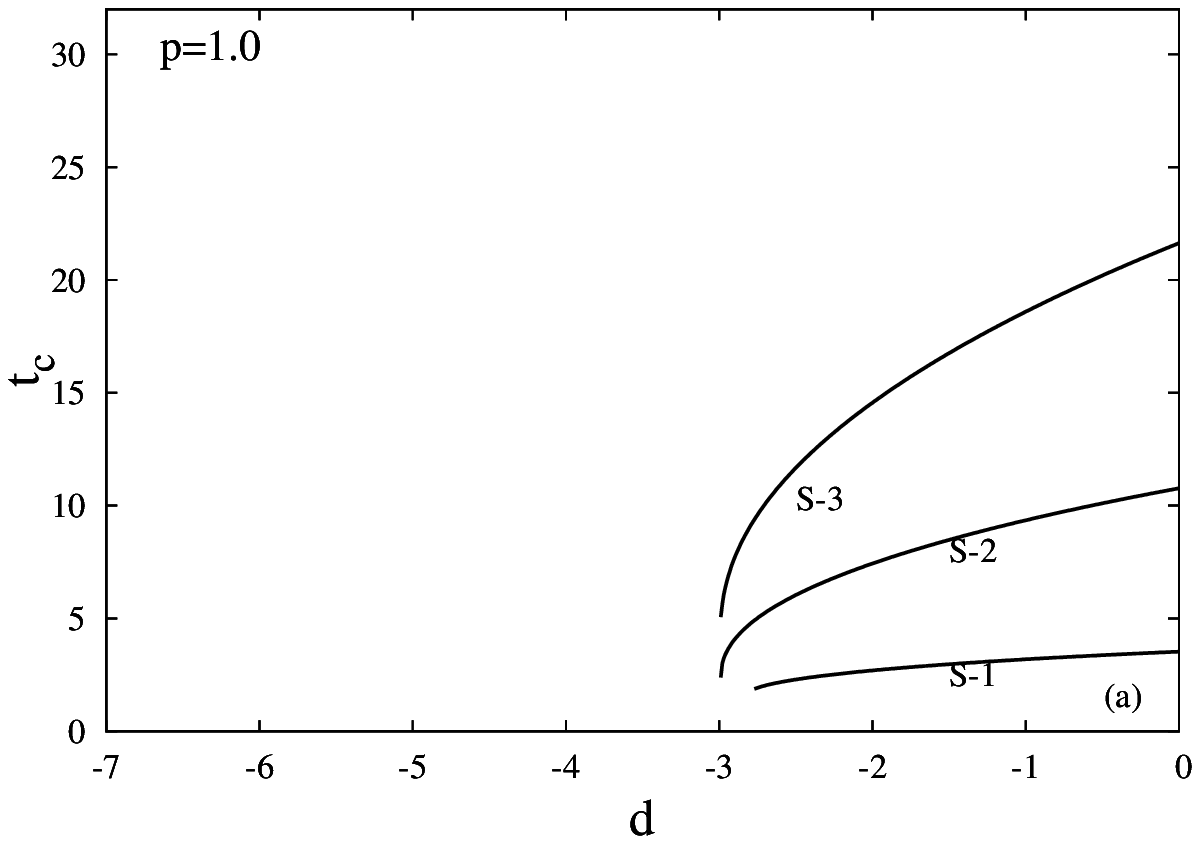, width=7cm}
\epsfig{file=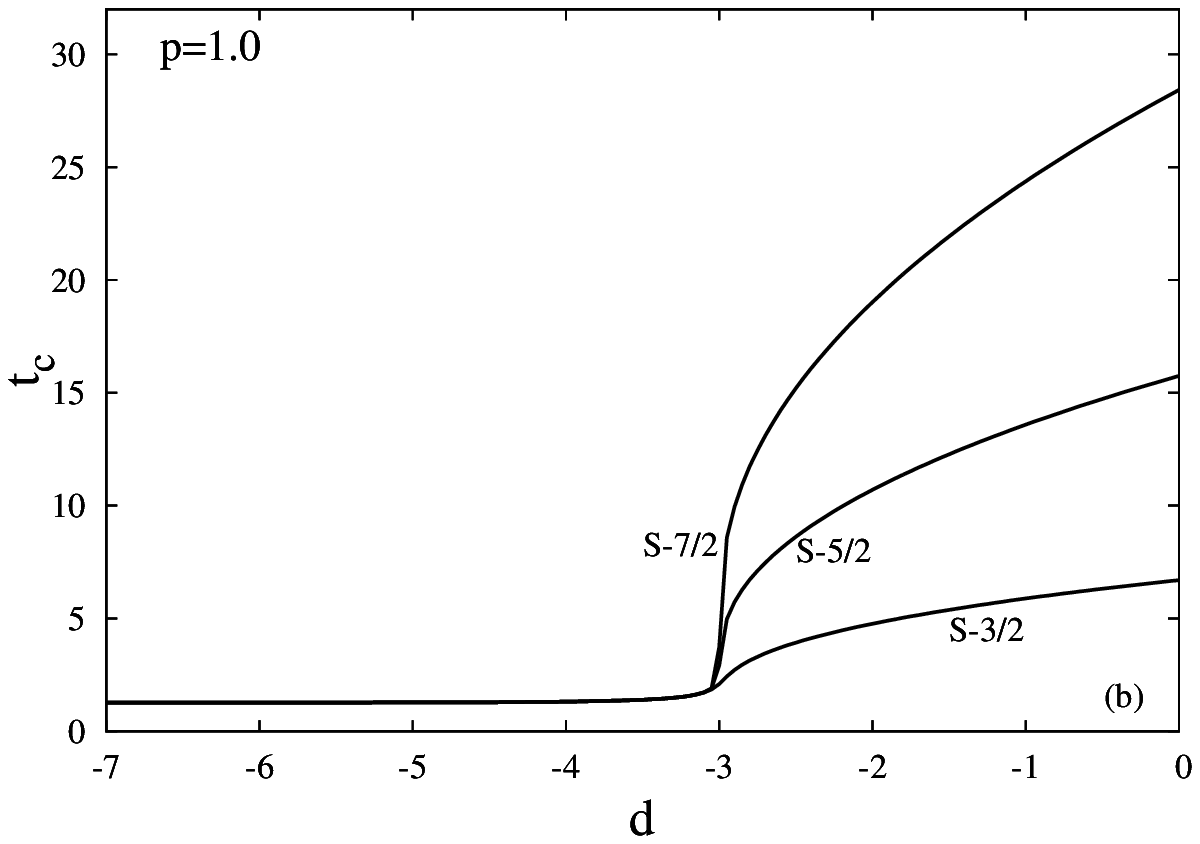, width=7cm}

\epsfig{file=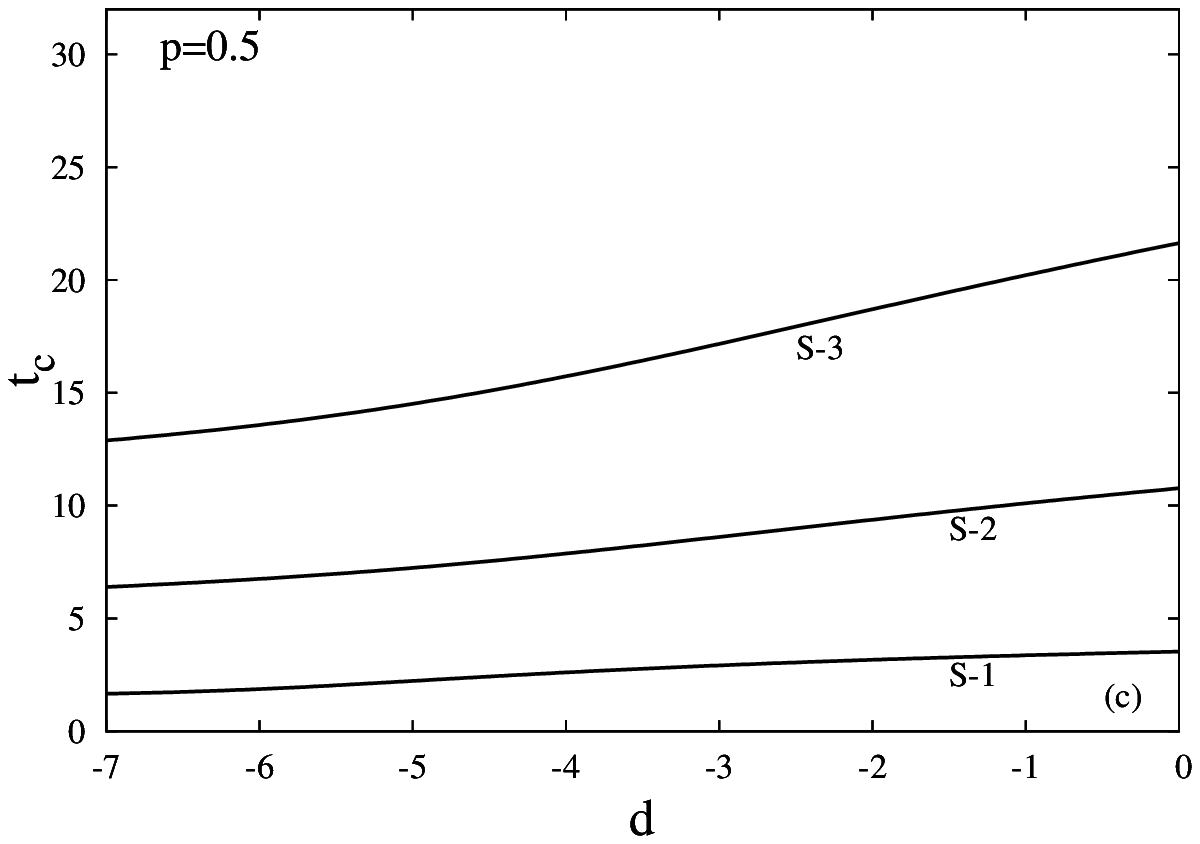, width=7cm}
\epsfig{file=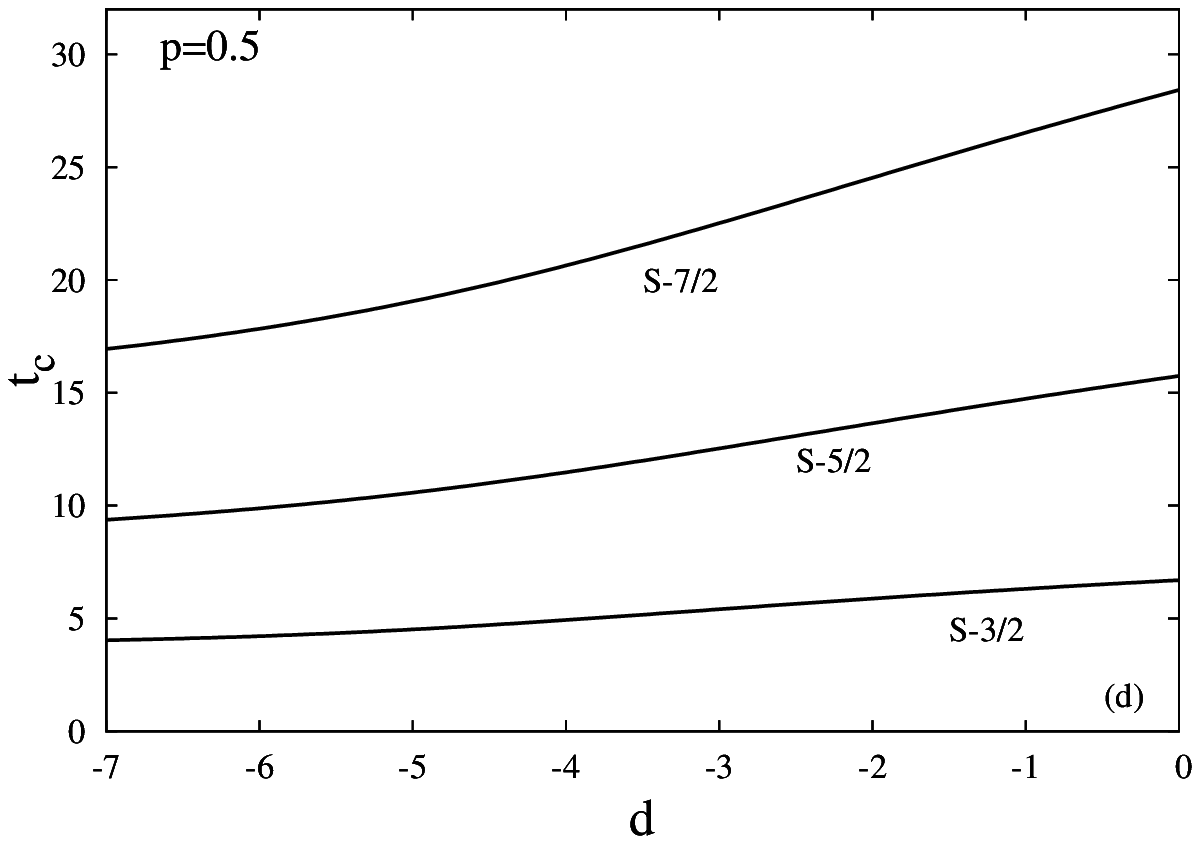, width=7cm}

\epsfig{file=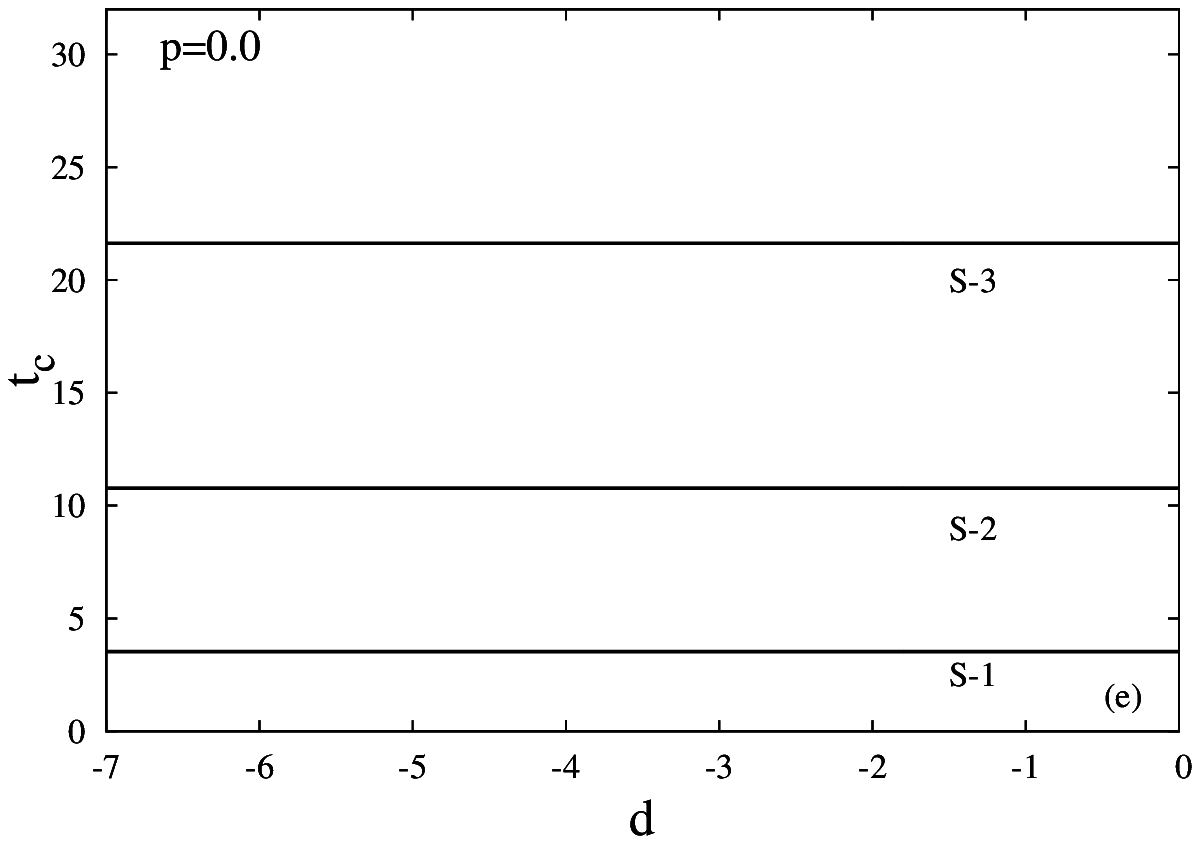, width=7cm}
\epsfig{file=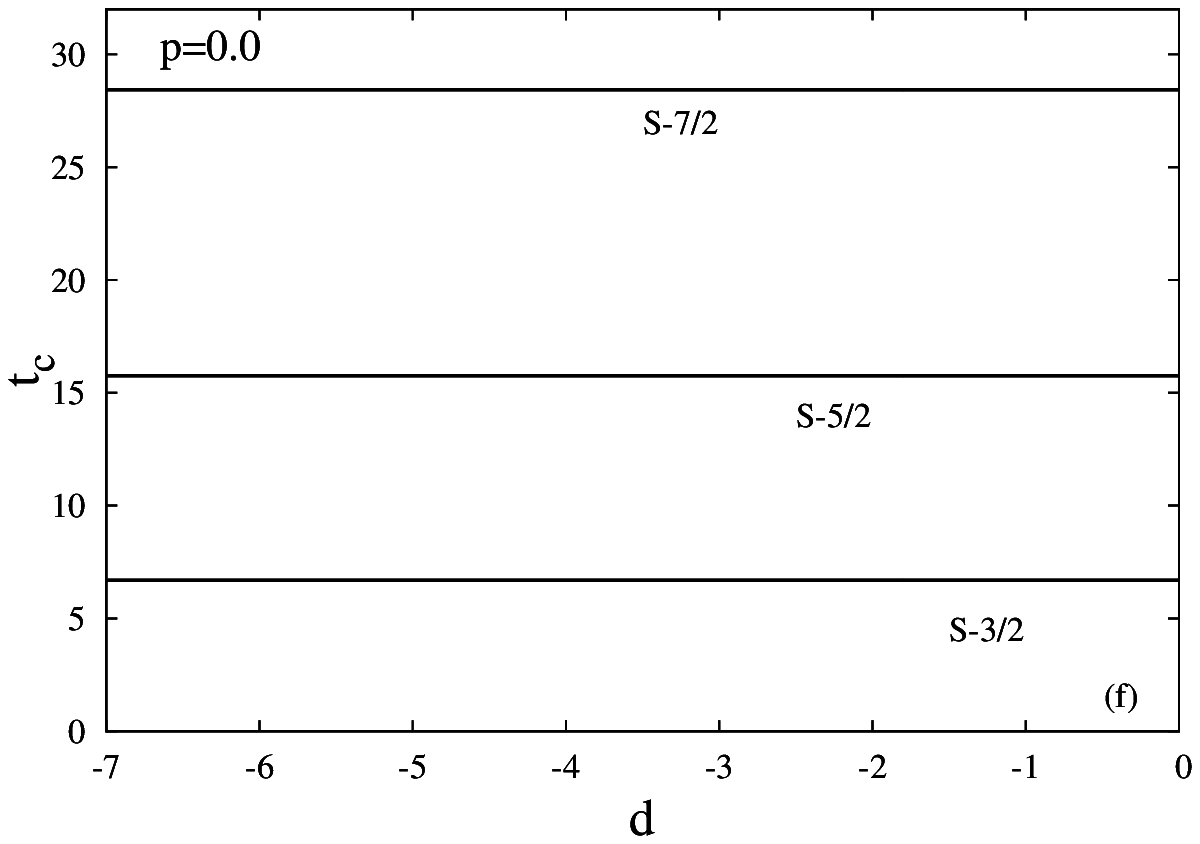, width=7cm}
\caption{Variation of the critical temperature of the spin-S BC 
model with crystal field for selected values of $p$.}\label{sek1}
\end{figure}

The phase diagrams of the spin-S BC model for various values of $p$ in $(d-t)$ plane can be seen in Figs. 
\ref{sek1} (a), (c), (e) for integer S and (b), (d), (f) for half integer S. Indeed these diagrams have already been obtained for specific values of $S$ in the literature with several methods. For instance, for $S-1,3/2,2$ within the PA \cite{ref34},  
$S-1$ within MC \cite{ref5}, FCA \cite{ref9}, MFA \cite{ref12,ref13,ref14,ref15} and EFT \cite{ref18,ref19,ref20},
$S-3/2$ model with crystal field dilution EFT \cite{ref30}, $S-2$ EFT \cite{ref33}.

As seen in Fig. \ref{sek1}, the phase diagrams of both the integer (Fig. \ref{sek1} (a),(c),(e)) and half integer (Fig. \ref{sek1} (b),(d),(f)) model evolve into parallel lines respect to the $d$ axis, as $p$ decreases. These parallel lines has $t_c$ values of critical temperature of the simple cubic lattice at $d=0$ of the related spin-S Ising model. The main qualitative difference between the evolution of the phase diagrams of integer and half integer spin model is that, decreasing $p$ creates ordered phase in the large negative valued $d$ region for integer model, which is absent for $p=1$ (i.e. homogenously distributed crystal field).  Note that, tricritical point of the integer model at $p=1$ disappears when $p$ decreases (compare Fig. \ref{sek1} (a) and (c)).  

This evolution has a simple physical mechanism. For $p=1$, integer spin model has ground state that all spins occuppied in the state $s=0$. This makes the expectation value of the internal energy minimum for low temperature. When $p$ decreases, some sites tends to be influenced by  the crystal field $d=0$. Note that for $d=0$, $s=\pm S$ is the state that makes the energy minimum (for low temperature) for the spin-S model. This means that as $p$ decreases, some spins start to fill $s=\pm S$ states. Exchange couplings between the nearest neighbor spins tend to align the nearest neighbor spins parallel with each other. When the value of $p$ becomes lower than a specific $p$, then the ground state becomes ordered. 

\begin{figure}[h]
\epsfig{file=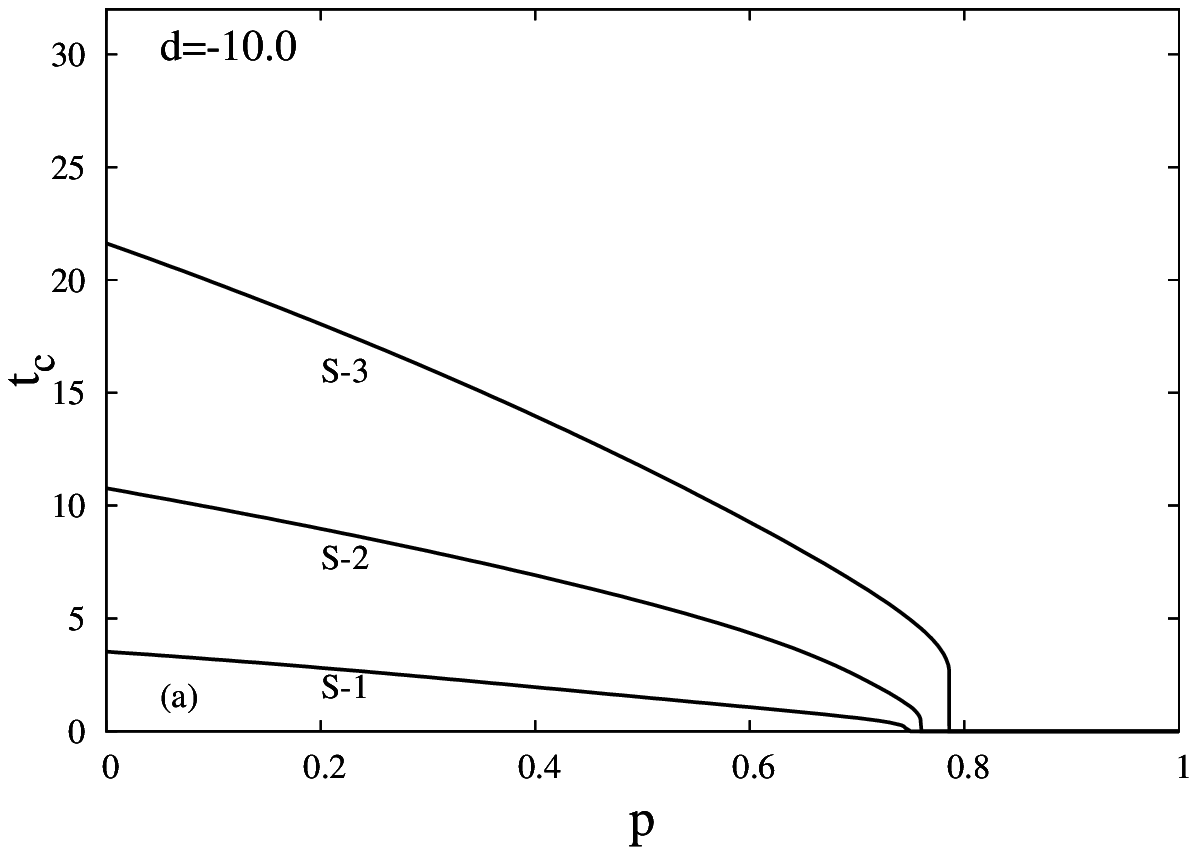, width=7cm}
\epsfig{file=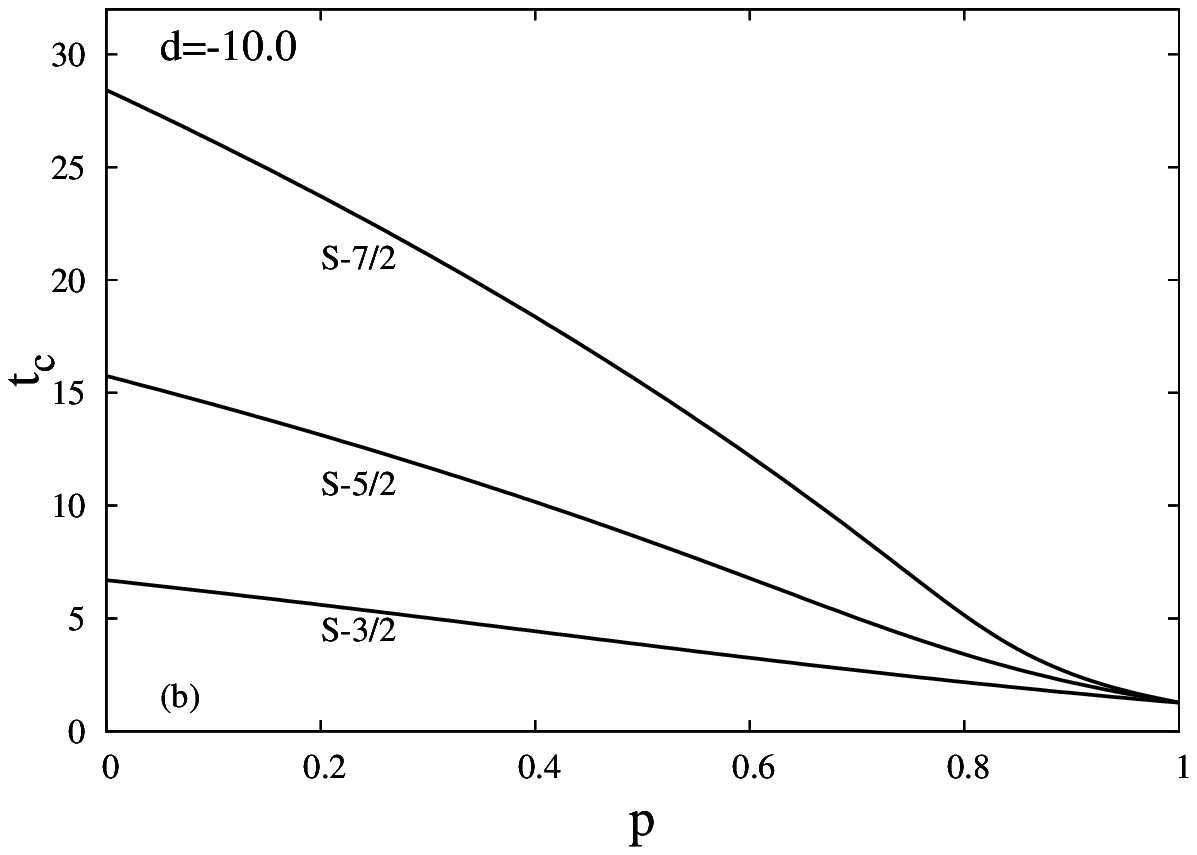, width=7cm}
\caption{Variation of the critical temperature of the spin-S BC 
model with $p$ for the value of  $d=-10.0$.}\label{sek2}
\end{figure}

Since we want to focus on the hysteresis behaviors of the large negative valued crystal field region, we depict the phase diagrams at the crystal field value of $d=-10.0$, in $(t_c,p)$ plane for integer (Fig. \ref{sek2} (a)) and half integer (Fig. \ref{sek2} (b)) spin-S BC model. Above mentioned difference can be seen in these diagrams clearly. As seen in Fig. \re{sek2} (b) for half integer model, the ground state is ordered, regardless the value of $p$, while this is not  the case for the integer model as seen in  Fig. \re{sek2} (a). Decreasing $p$ (diluting the crystal field) results in an ordered ground state for the integer model.

\subsection{Hysteresis behaviors}

We mainly focus on the hysteresis behavior of the spin-S BC model in  the large negative values of $d$ and low temperature. Let us call this region as limit region. In general, as explained in Ref. \cite{ref47}, for the limit region,  spin-S BC model could exhibit $2S$-windowed hysteresis loops. As temperature rises, hysteresis loop of integer spin-S BC model evolves as $2S\rightarrow 2S-2 \rightarrow \ldots\rightarrow 2\rightarrow 0$, while this sequence for half integer 
spin-S BC model is $2S\rightarrow 2S-1 \rightarrow 2S-3 \rightarrow 
\ldots\rightarrow 2\rightarrow 0$. Hysteresis loop that has $0$ number of 
windows means that, there is no history dependent variation of the order 
parameter on the longitudinal magnetic field. The physical mechanisms behind  this behavior were explained in Ref. \cite{ref47}, in detail.

On the other hand, crystal field diluted $S-1$ BC model has double, triple or single  hysteresis 
behavior according to the value of $p$ and other Hamiltonian parameters \cite{ref40}. Thus it is expected that, integer spin-S BC model could display similar behavior, i.e. dilution of crystal field may cause $2S+1$ windowed hysteresis loop for integer spin-S BC model, in limit region.

As an example of the variation of the hysteresis loops with crystal field dilution, we depict the hysteresis loops of $S-2$ and $S-5/2$ BC model in Fig. \re{sek3}. As explained in Ref. \cite{ref47}, spin-S BC model has $2S$ windowed loop, which can be seen in Fig. \re{sek3} (a) and (b). When the crystal field is diluted, integer model can produce central loop (which is the sign of the ferromagnetic phase, see Fig. \re{sek3} (c)) and the number of windows of the loop decreases for both integer (compare Figs. \re{sek3} (a) and (c)) and half integer model (compare Figs. \re{sek3} (b) and (d)). When the temperature rises from this crystal field diluted system, loops become one central windowed (see Figs. \re{sek3} (e) and (f)).

\begin{figure}[h]
\epsfig{file=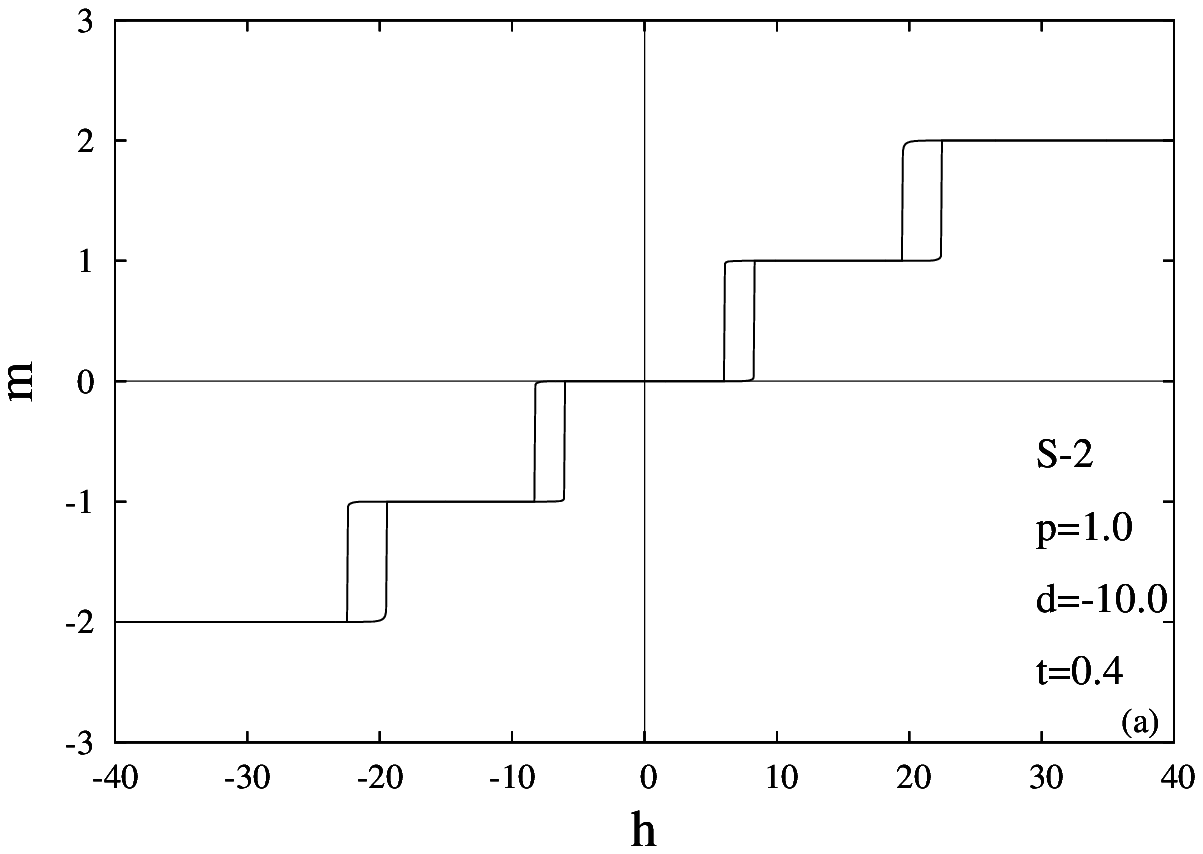, width=7cm}
\epsfig{file=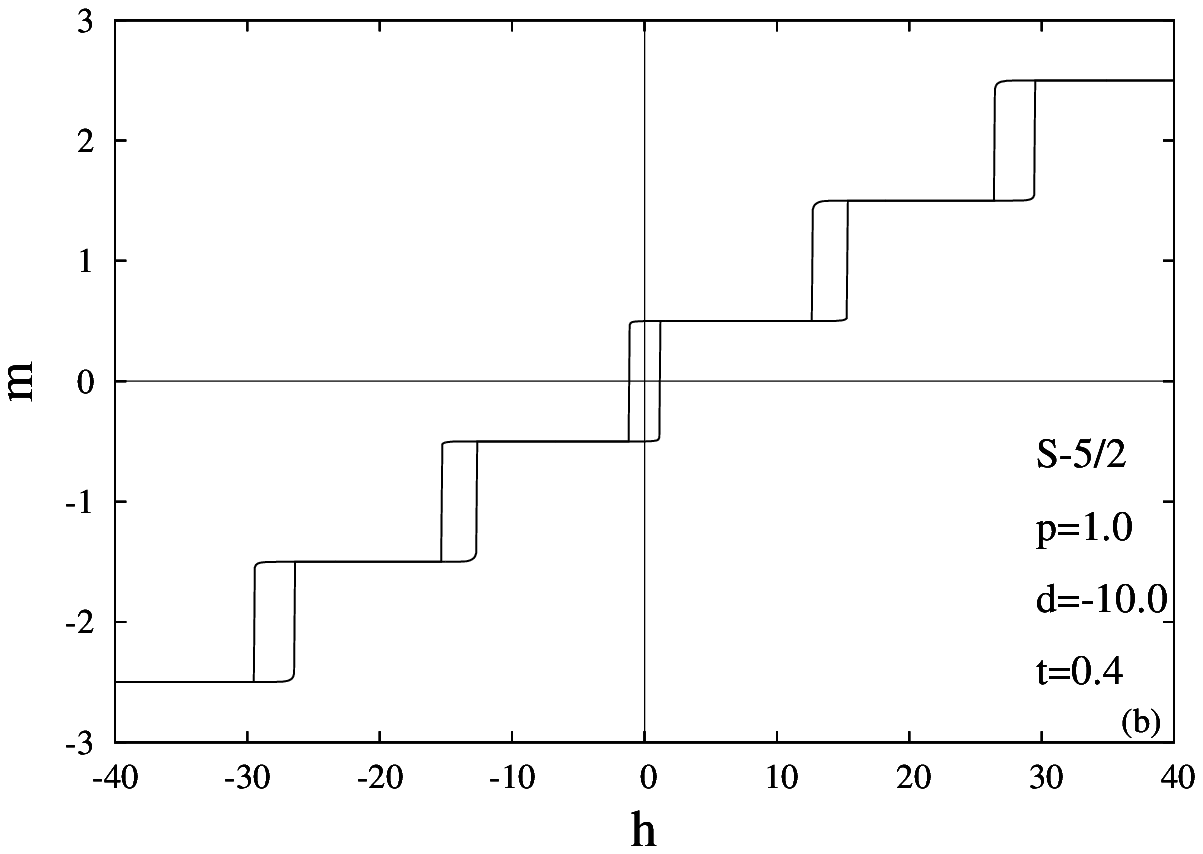, width=7cm}

\epsfig{file=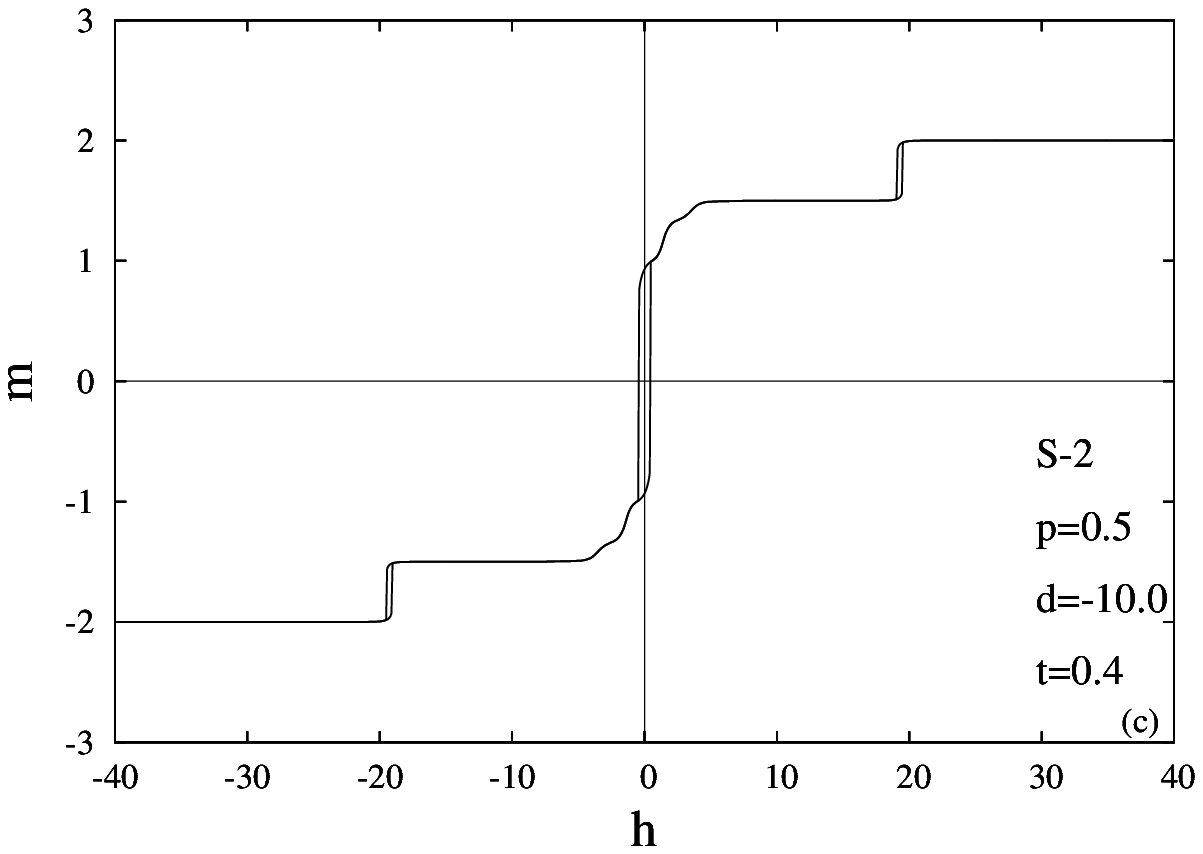, width=7cm}
\epsfig{file=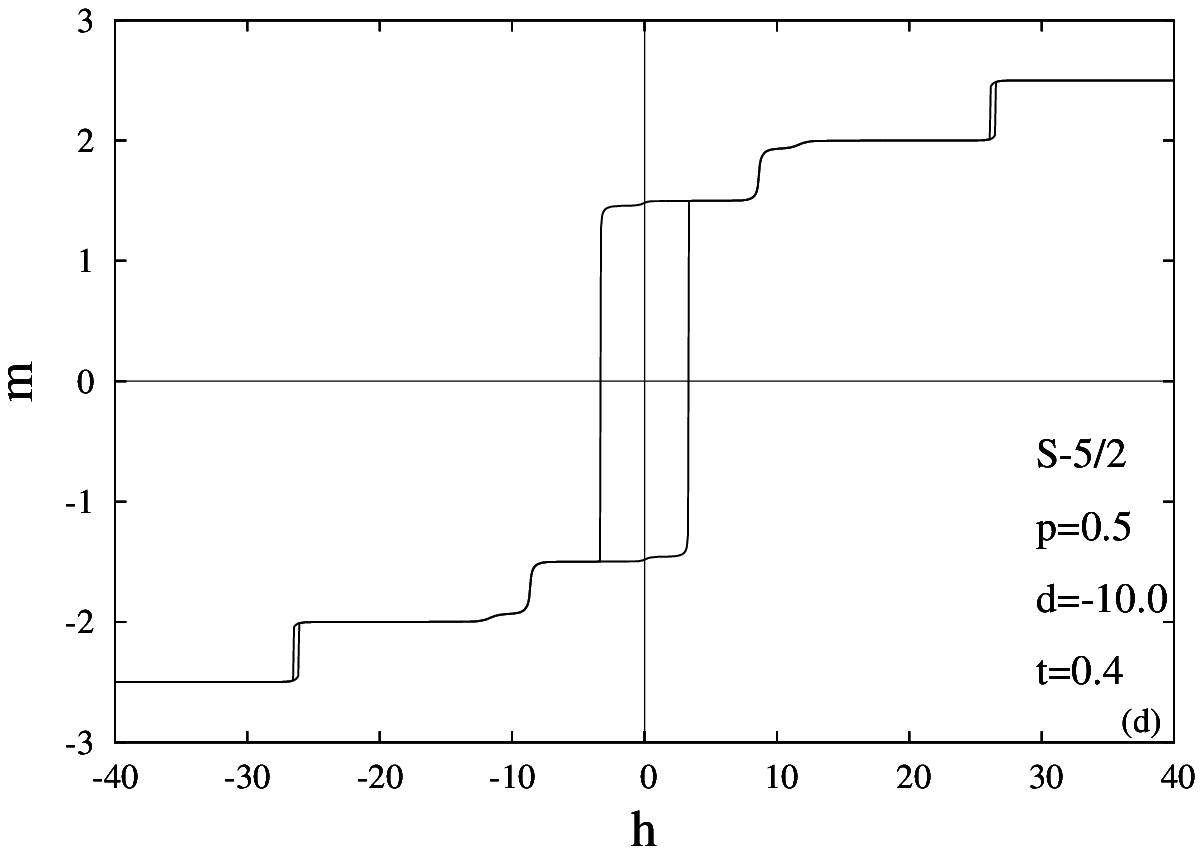, width=7cm}

\epsfig{file=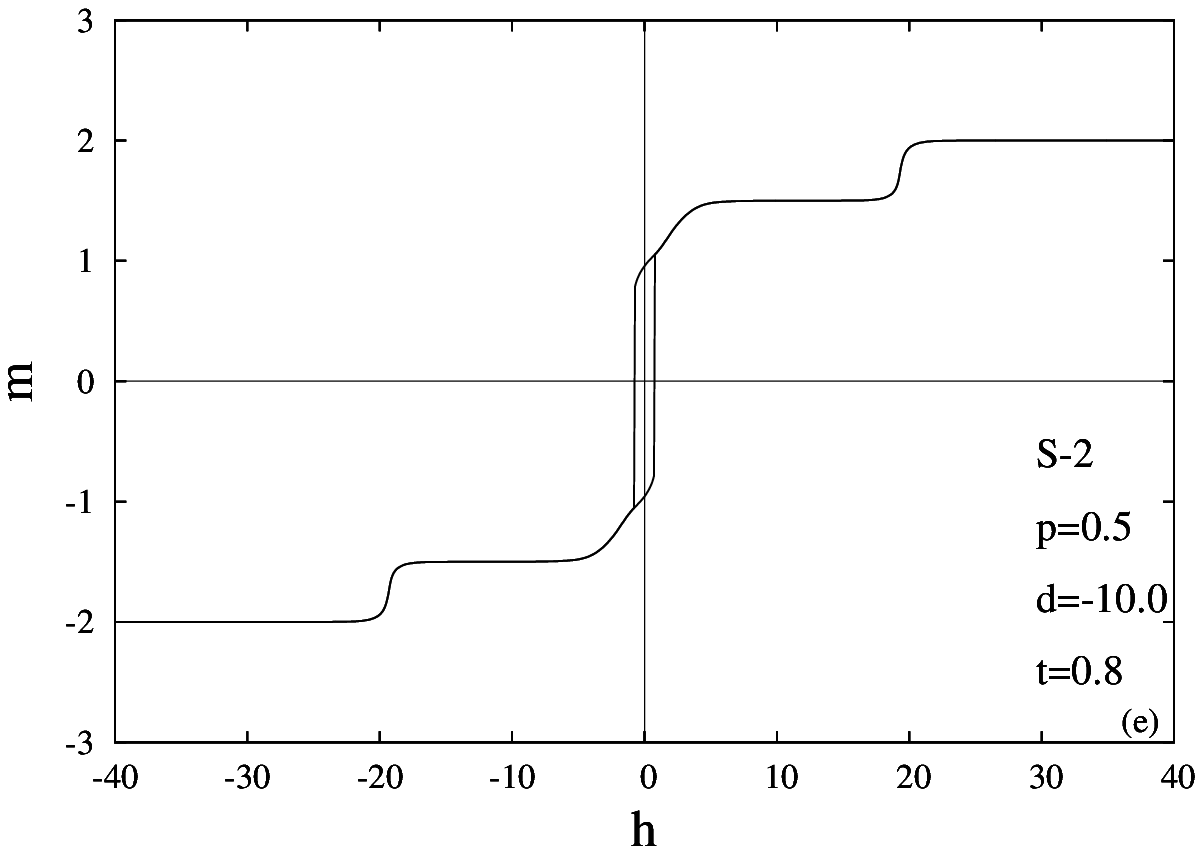, width=7cm}
\epsfig{file=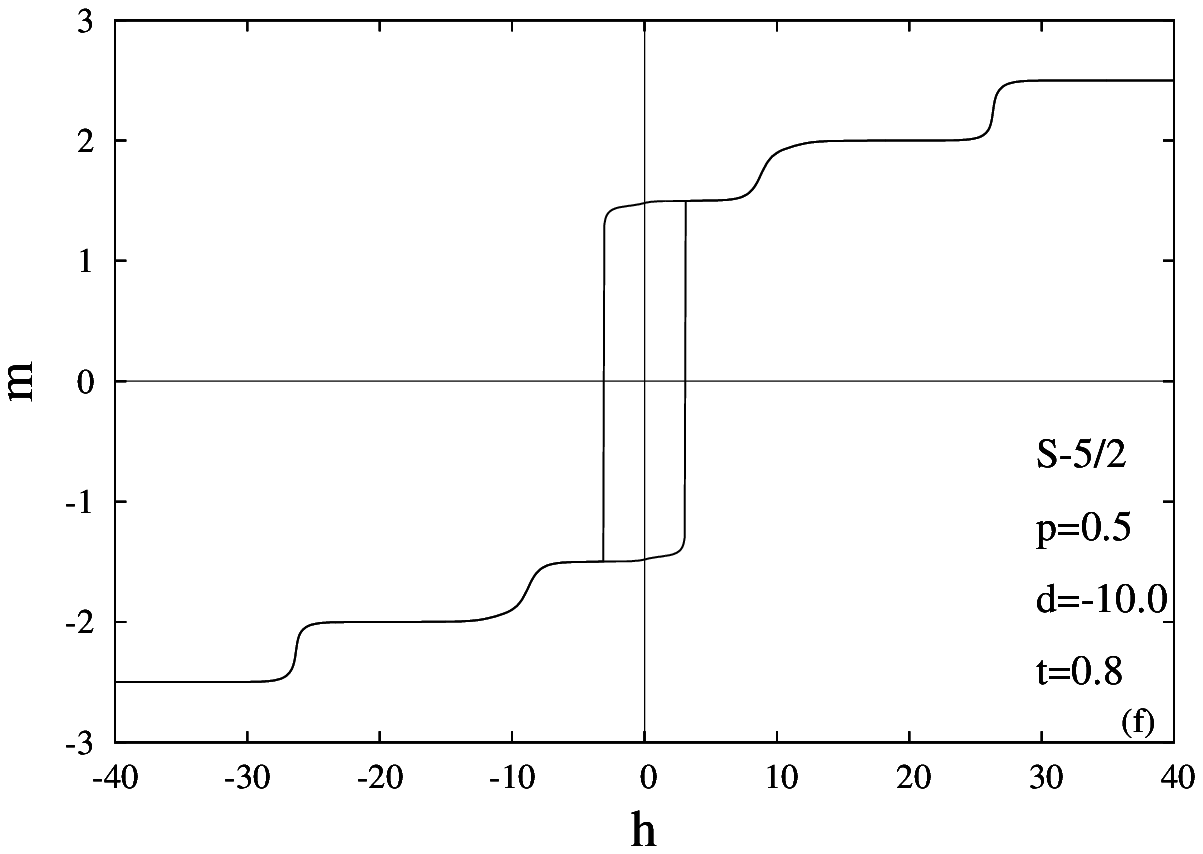, width=7cm}
\caption{Hysteresis loops of $S-2$ and $S-5/2$ BC models for various values 
of $p,d$ and $t$.}\label{sek3}
\end{figure}

By inspection, we can determine the number $n$, which is the number of windows appearing in hysteresis loop. The regions in $(p,t)$ plane which have different $n$ can be seen in In Fig. \re{sek4}. These figures  summarize the hysteresis characteristics in limit region for spin-S BC model. 
In general borders between the regions that have different $n$ are lines, except the critical line (which are shown as dotted line in Fig. \re{sek4}). Critical line seperates the ordered phase and disordered phase, and these lines are nothing but the lines depicted in Fig. \re{sek2}. This seperation shows itself in the characteristics of the hysteresis loops. The loops that belong to ordered phase have odd numbered windows, on the other hand hysteresis loops of the disordered phase have even numbered windows.

In a limit region, the variation of hysteresis loops with rising dilution (i.e. decreasing $p$) of half integer spin-S BC model is given by a sequence $2S\rightarrow 2S-2\rightarrow\ldots \rightarrow 1$. This fact can be seen from Figs. \re{sek4} (d), (e) and (f). This sequence is more complicated for the integer spin model, as we can see from Figs. \re{sek4} (a), (b) and (c). From Figs. \re{sek4} (a), (b)  we can see the variation of the hysteresis loop of the integer spin model in limit region  given as a sequence $2S\rightarrow 2S+1\rightarrow 2S-1 \rightarrow 2S-3  \ldots \rightarrow 1$. But this is not in the case for S-3 BC model. As seen in Fig. \re{sek4} (c), the sequence of this model is $2S\rightarrow 2S-2\rightarrow 2S-1 \rightarrow 2S-3  \ldots \rightarrow 1$. When the rising dilution results in a transition from the disordered phase to the ordered one, number of windows in the hysteresis loop is increases by one. This increment comes from the appearance of the central loop, which is typical for ordered phase. This fact can also  be seen by comparing  Figs. \re{sek3} (a) with (c). To conclude, we can say that the number of windows of the hysteresis loop of half integer spin BC model regularly decreases by two, while crystal field is diluted. On the ther hand same behavior is valid for integer spin BC model except one point. The number of windows of the hysrteresis loop increase by one, if dilution of the crystal field give rise to the phase transition from the disordered phase to the ordered phase.  This  declining behavior ends with hysteresis loop with one window, both for integer and half integer spin BC  model. 

\begin{figure}[h]
\epsfig{file=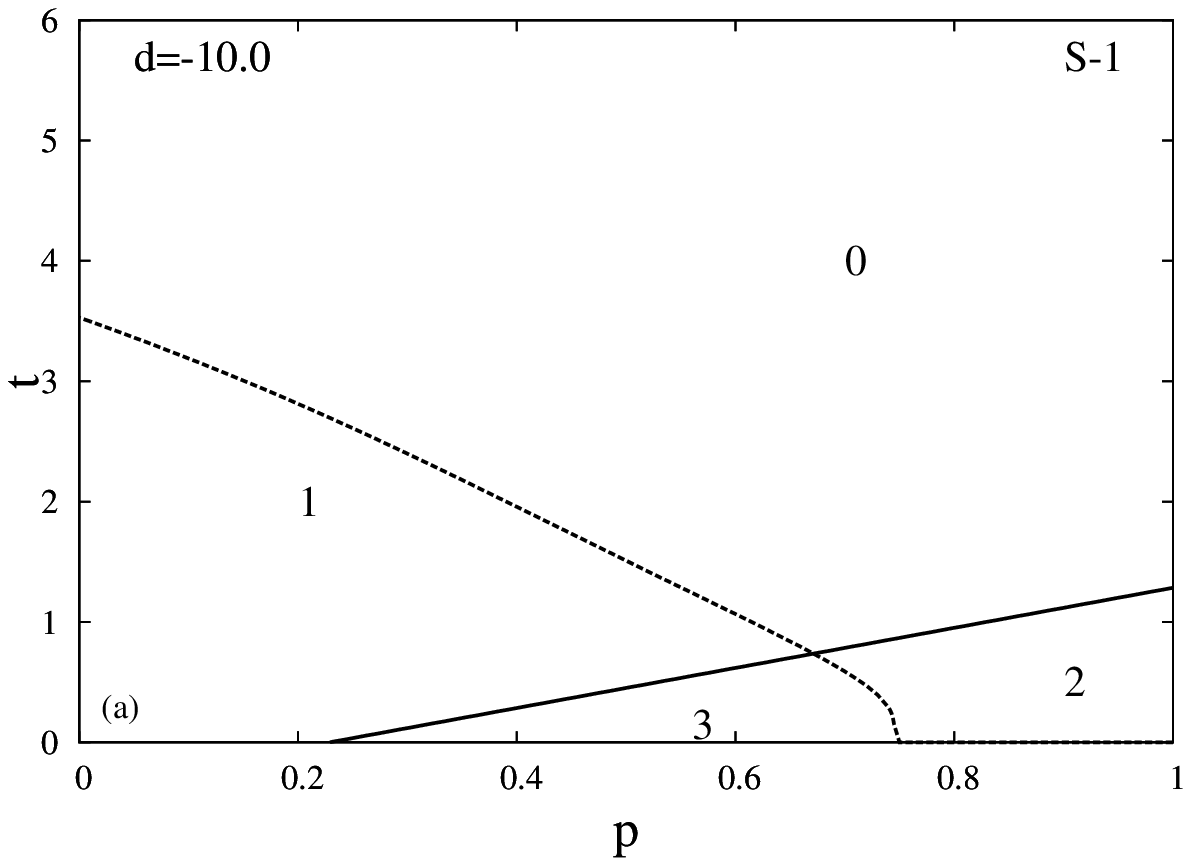, width=7cm}
\epsfig{file=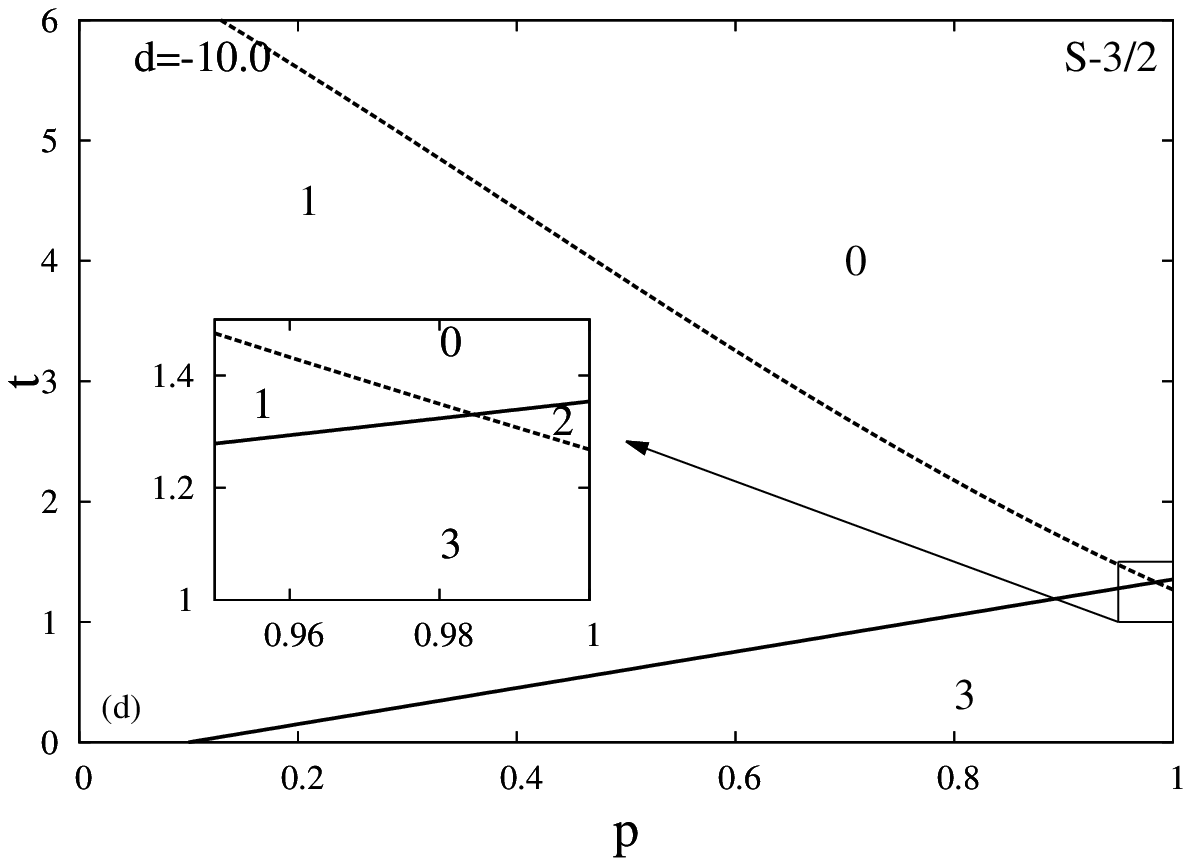, width=7cm}

\epsfig{file=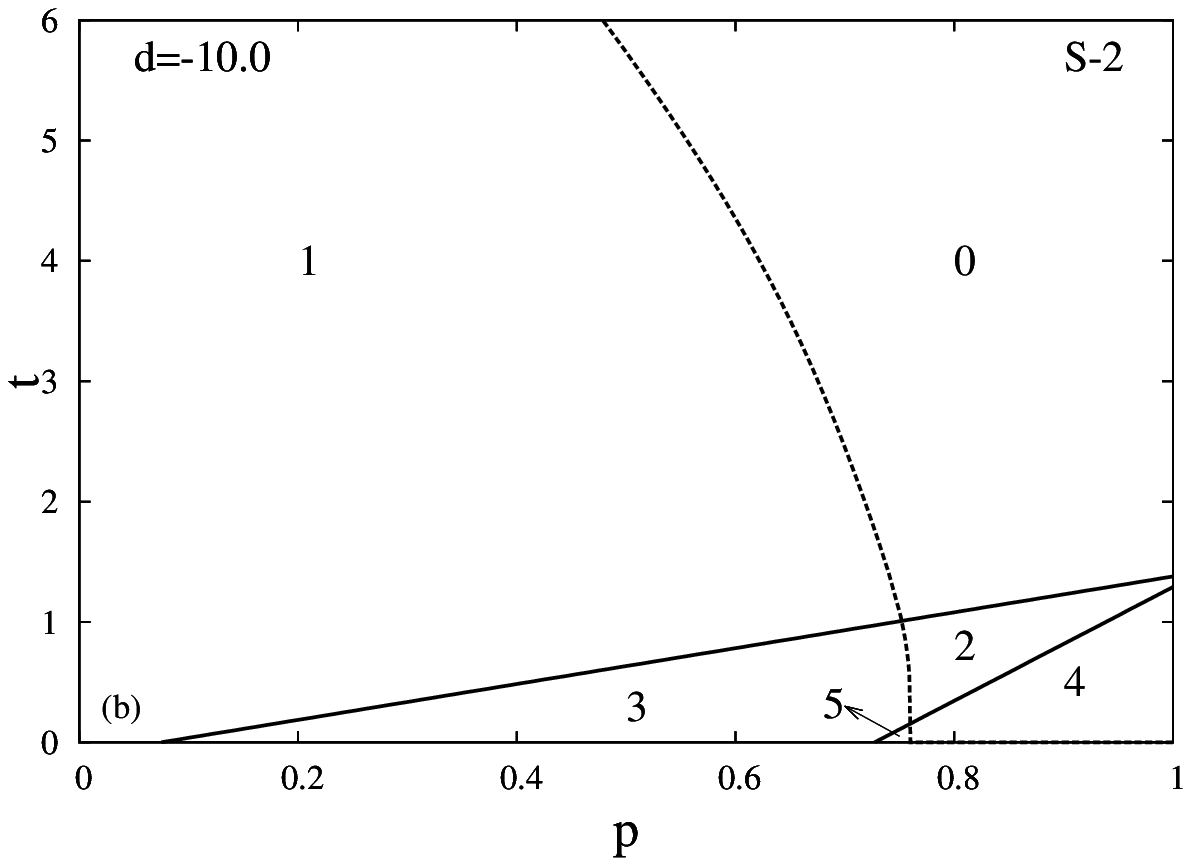, width=7cm}
\epsfig{file=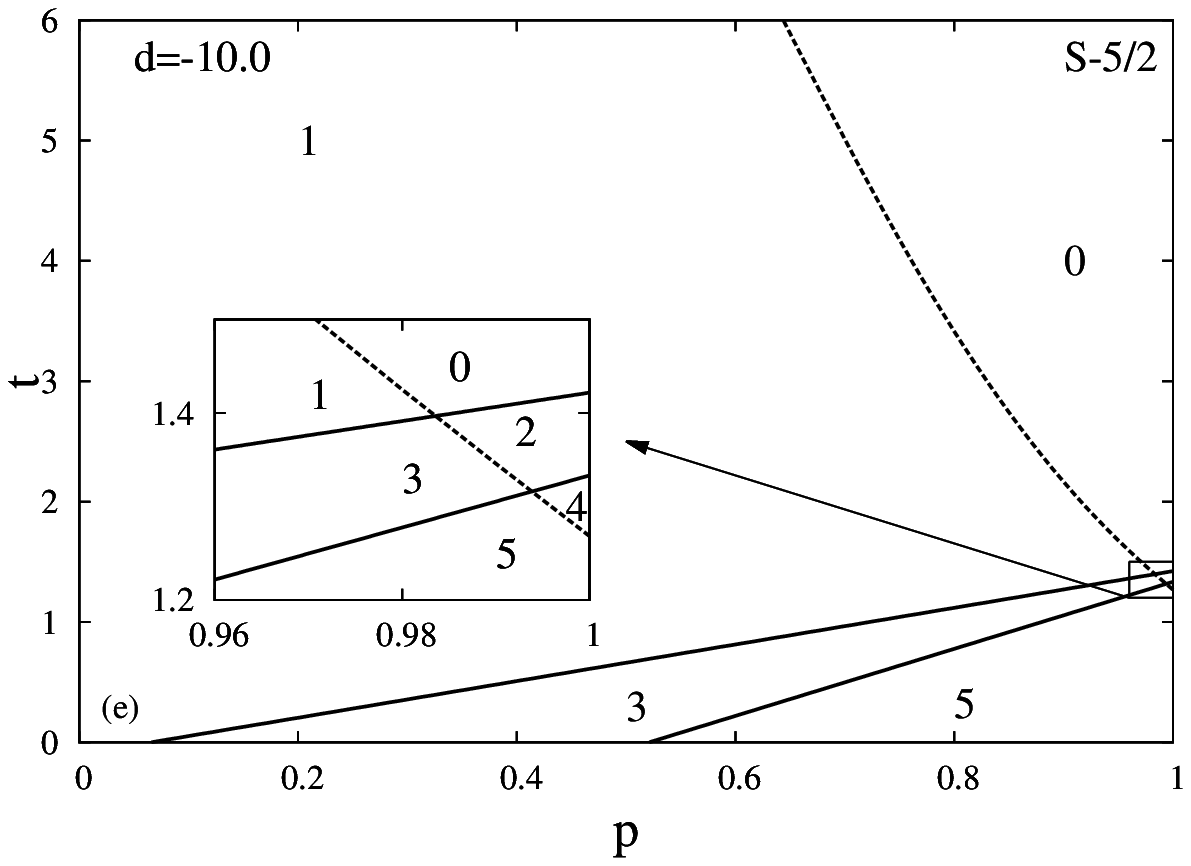, width=7cm}

\epsfig{file=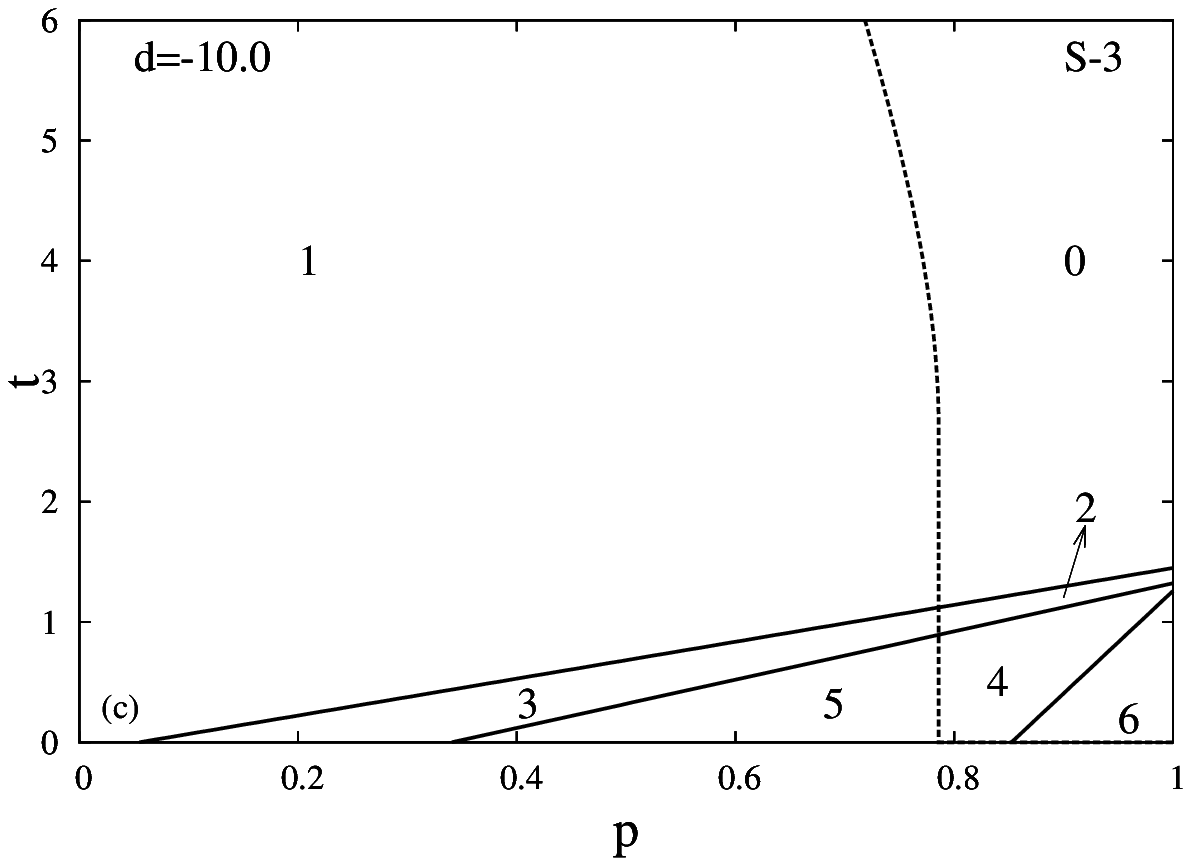, width=7cm}
\epsfig{file=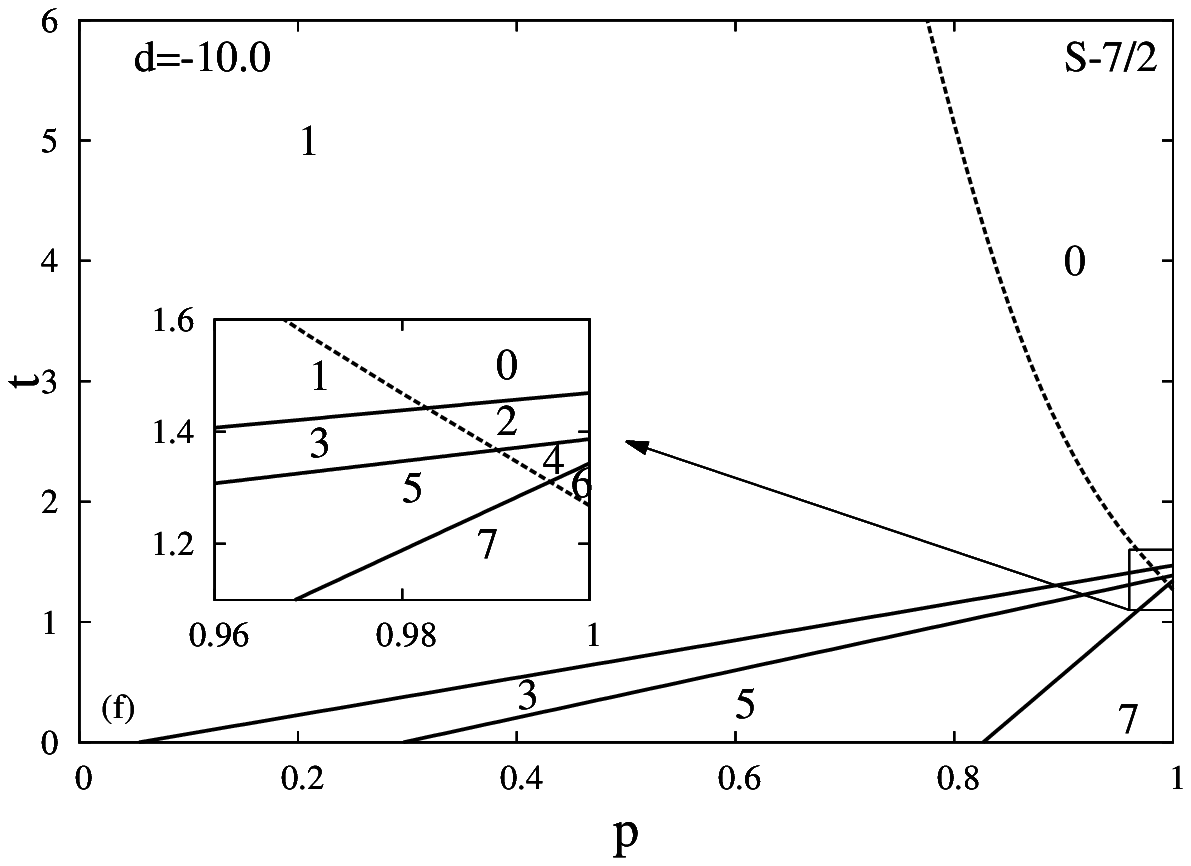, width=7cm}
\caption{The regions that have n-windowed hysteresis loops in $(p,t)$ plane for $S-2$ and $S-5/2$ BC model. n is mentioned in each region. The  dotted lines represent critical lines.}\label{sek4}
\end{figure}

These interesting behaviors can be detectable by means of the experimental methods via the magnetic susceptibility, which is defined by $\chi=\partial m/ \partial h$. Sharp change of the magnetization with applied magnetic field causes a  sharp peak in magnetic susceptibility. Naturally, this does not directly point out the window of hysteresis loop. Instead, this signals the sharp change of the magnetization with the magnetic field. But this sharp change is generally history dependent,  that is, one window of the hysteresis loop. In order to clarify this point we depict the hysteresis loops and corresponding variations of the susceptibilities with magnetic field in Fig. \re{sek5}, for S-3 BC model with $d=-10.0$. As seen in Fig. \re{sek5} sharp peaks of the magnetic susceptibility correspond to the history dependent change of magnetization with the magnetic field (see Fig. \re{sek5} (a) and central peaks of  Figs. \re{sek5} (c) and (d)). Besides, as seen in Fig. \re{sek5} (b) all peaks of the magnetic susceptibilities are smooth and correspond to the smooth and history independent change of the magnetization with the magnetic field.

\begin{figure}[h]
\epsfig{file=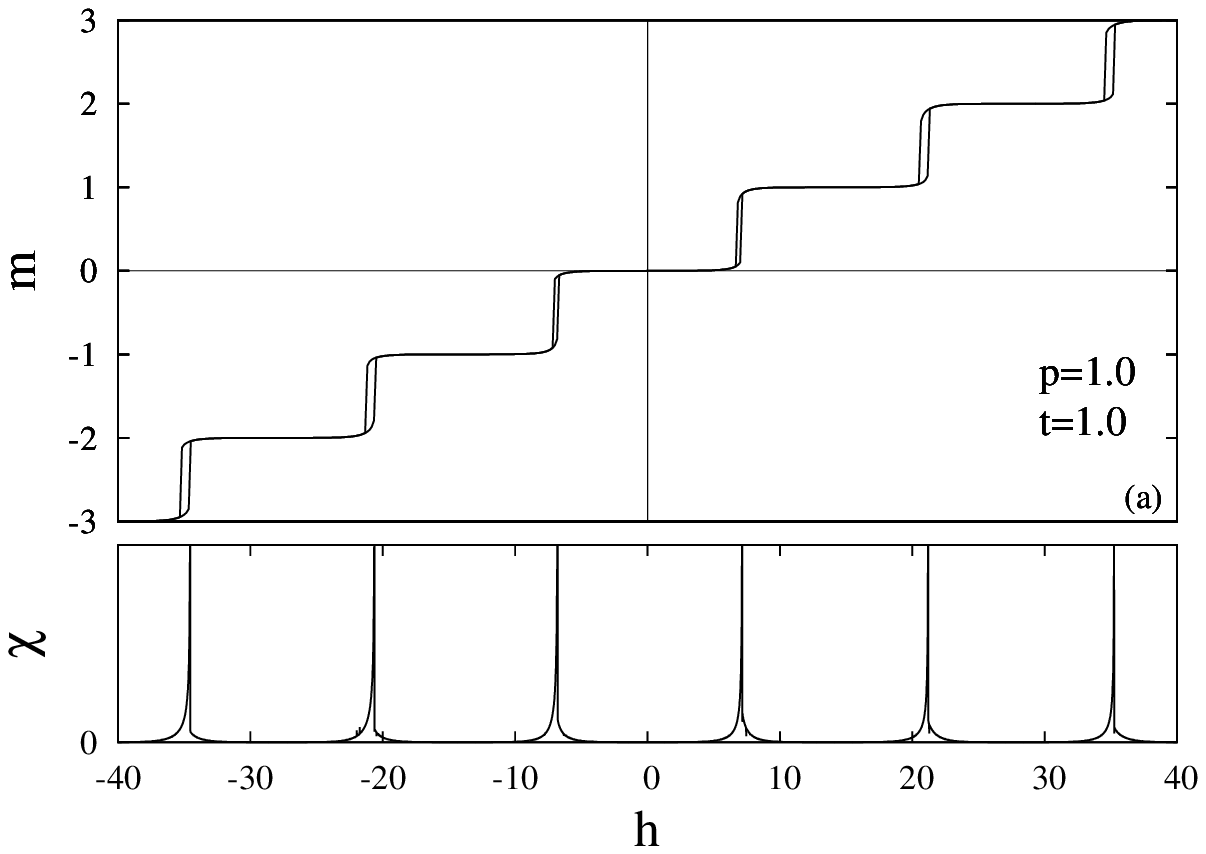, width=7cm}
\epsfig{file=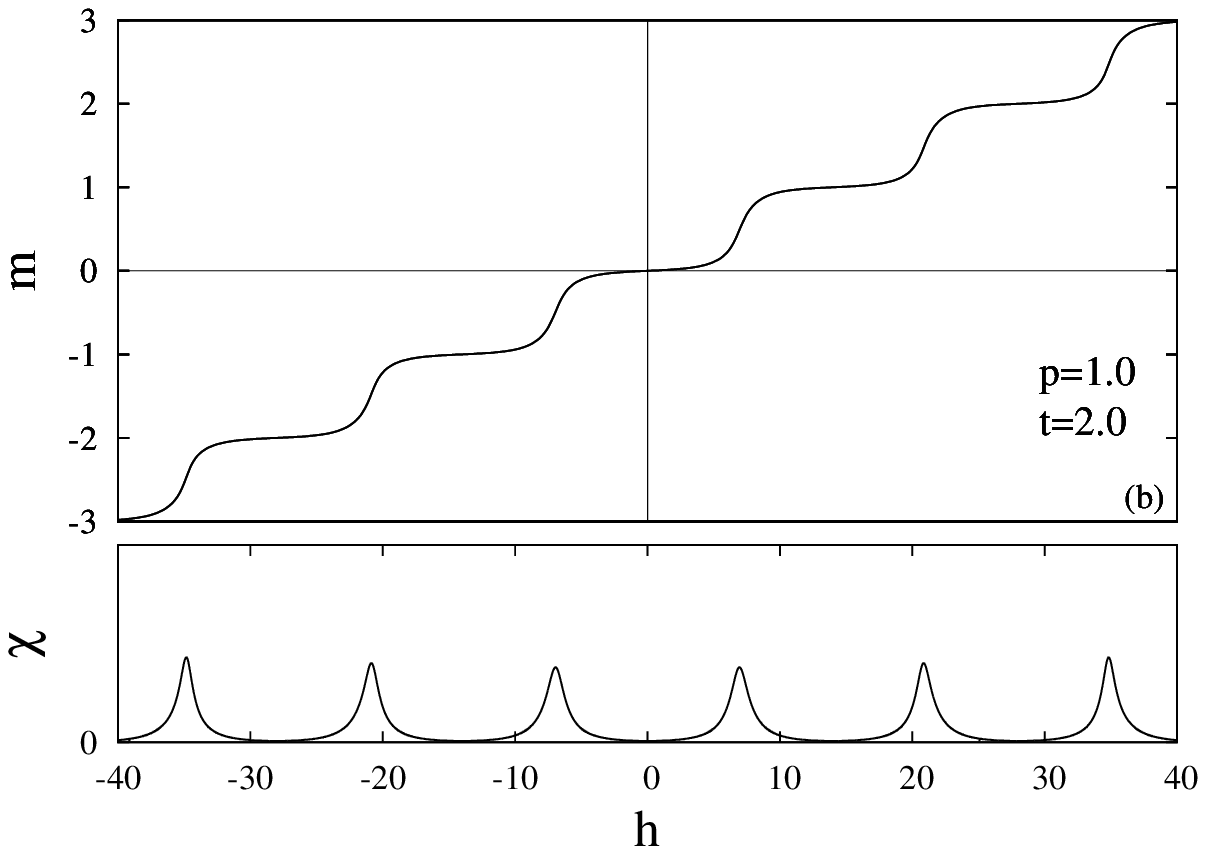, width=7cm}

\epsfig{file=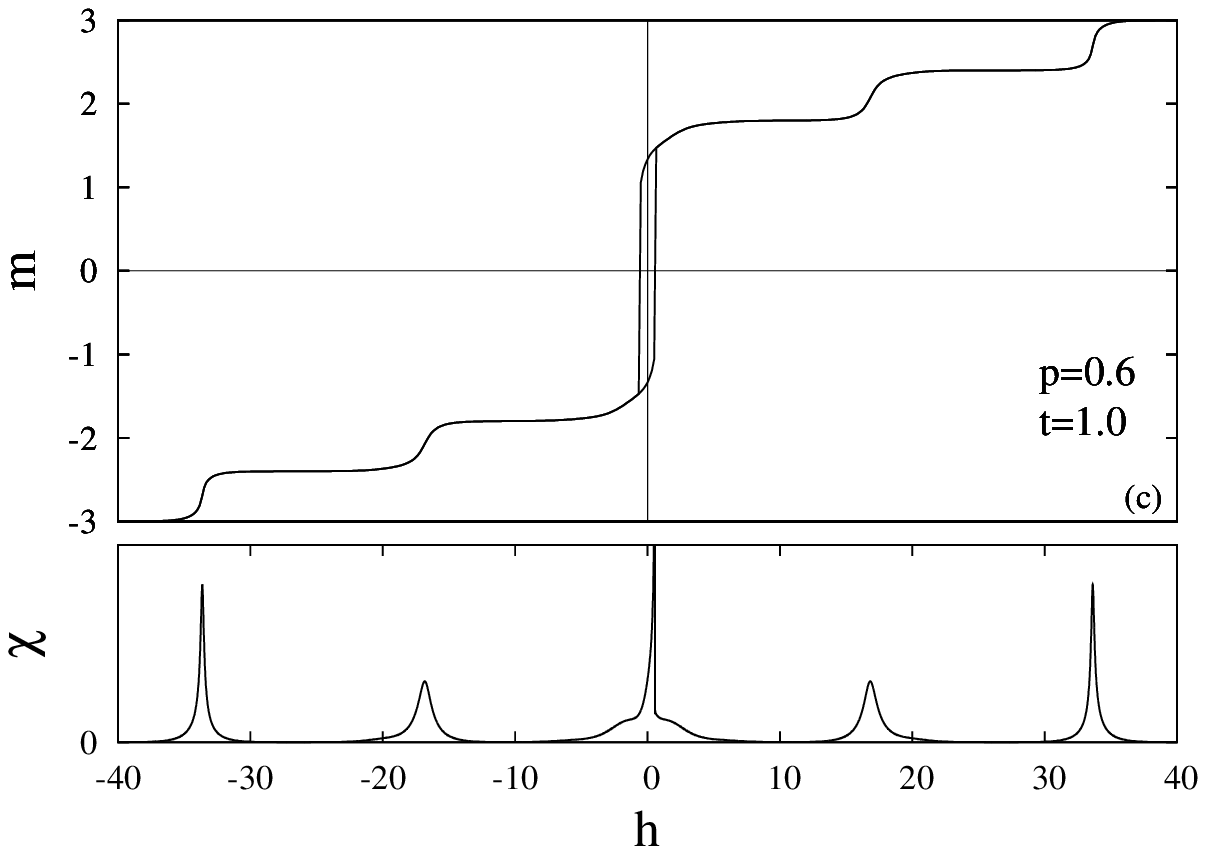, width=7cm}
\epsfig{file=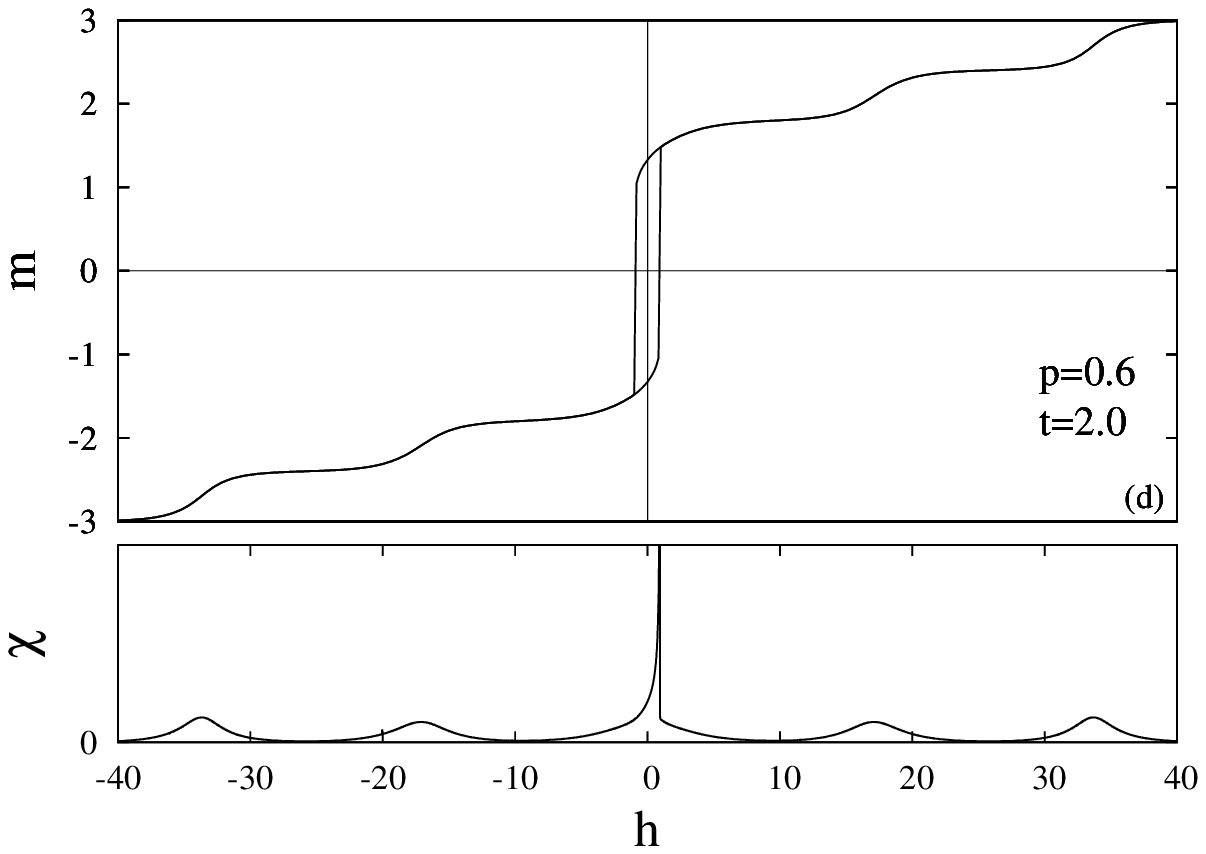, width=7cm}
\caption{Hysteresis loops and variation of magnetic susceptibility with magnetic field for S-3 BC model with crystal field value $d=-10.0$. The values of other Hamiltonian parameters are given in each figure. }\label{sek5}
\end{figure}

The physical mechanism behind these hysteresis behaviors is simple. As explained in Ref. \cite{ref47}, in limit region, most of the $s=0$ states are occupied by the spins for integer spin BC model, and most of the spins are in a state $s=1/2$ for half integer BC model. Magnetic field can induce transitions from these states to the higher $s$ states. For instance, rising $H$ in limit region, can induce $0\rightarrow 1$ transition for integer spin model and  $1/2\rightarrow 3/2$ transition for the half integer spin model. Further rising magnetic field can induce  $1\rightarrow 2$ ($3/2\rightarrow 5/2$) transition for integer (half integer) spin BC model, respectively. This plateau behavior of magnetization and inducing history dependent transitions between these plateaus in limit region can be seen in Fig. \re{sek3} (a) for integer spin model  and  (b) for half integer spin model. As dilution in case (i.e. lowering of the value of $p$), $(1-p)$ ratio of the lattice sites are influence of the zero crystal field. This means that, spins belonging these sites are in states $s=\pm S$ for spin-S BC model at limit region, if we do not consider the spin-spin interaction. The competition between the spin-spin interaction and the dilution says the last word. According to the value of $p$, plateau behavior mentioned for $p=1$ case disappears (compare Figs. \re{sek3} (a) with (c) and (b) with (d)). Rising temperature causes a shrink of the windows, can be seen by comparing  Figs. \re{sek3} (c) with (e) and (d) with (f).


\section{Conclusion}\label{conclusion}

Hysteresis characteristics of the crystal field diluted general Spin-S ($S>1$) BC model have 
been studied within the EFT. General results obtained about the system for low temperature and large negative values of the crystal field,  after systematic investigations.  This limit case is called as limit region in the text. This work can be considered as the generalization of Ref. 
\cite{ref47}, which was about the general spin-S BC model without any dilution of the crystal field. That work was concluded with, in general spin-S BC model could display $2S$ windowed hysteresis loops in limit region. 

In the limit region, it has been demonstrated that, rising dilution systematically decreases the number of windows of the hysreresis loop both for integer and half integer BC model. Except one point, which is when rising dilution causes the phase transition from then disordered phase to the ordered phase for integer S. In this case number of windows increases by one. It has also been  demonstrated that, rising dilution of the crystal field could develop central loop of the hysteresis when the phase transition occurs. The regions  that have different number of windows (of the hysteresis loop)  have been given for some selected values of S.

Besides, the relation between this multiple hysteresis behavior and the magnetic susceptibility of the system has also been constructed. This point is important to determine these multiple hysteresis behaviors experimentally. These results need to be generalized to anisotropic Heisenberg model, and this will be our 
next work. We hope that the results
obtained in this work may be beneficial form both theoretical and
experimental points of view.

\end{document}